\documentstyle[times,pramana,epsf,floats]{ias}
\def\g{\gamma}
\def\G{\Gamma}
\def\s{\sigma}
\begin{document}
\mark{{Identifying new physics contributions in the Higgs sector at
linear $e^+e^-$ colliders}{Santosh Kumar Rai}}
\title{Identifying new physics contributions in the Higgs sector at linear
$e^+e^-$ colliders}

\author{Santosh Kumar Rai \footnote{Talk presented at LCWS06,
Bangalore, India}}
\address{Harish-Chandra Research Institute, Chhatnag Road, Jhunsi,
Allahabad 211019, India}
\abstract{
Loop driven decay modes of the Higgs are sensitive to new physics
contributions because of new particles in the loops. To highlight this 
we look at the dilepton-dijet signal in the dominant Higgs production
channel at a linear $e^+ e^-$ collider. We show that by taking a 
simple ratio between cross-sections of two different final states such 
contributions can be very easily identified.}

\maketitle
\section{Introduction}
Higgs boson discovery will prove to be a crucial ingredient towards 
understanding the mechanism of electroweak symmetry breaking. 
Once discovered, a major goal would be
to determine its other intrinsic properties, couplings and its total 
width with high accuracy in a model independent way. The proposed 
future $e^+e^-$ linear colliders would be instrumental in achieving
very precise measurements of the Higgs boson properties.  

The partial width of the Higgs decaying to the massless gauge boson is
of special interest, since there are no tree level couplings of the
Higgs to them and any contribution is generated at the one-loop level.
The di-photon partial width gets contribution through massive charged
particles in the loops while the gluon-gluon partial width gets
contributions from the heavy quarks running in the loops. The effective
loop induced couplings of $H\g\g$ and $Hgg$ are sensitive to new
contributions from particles which appear in various extensions of the
SM. Not only do these decay modes provide for a possible probe of new
physics particles which are too heavy to be produced directly but they
are also sensitive to scales far beyond the Higgs mass. 
We take up the case of the enhancement in the partial decay width
of $H\to gg$ due to additional contributions coming from 
particles from theories of beyond SM (BSM) physics. 
Such additional heavy particles are predicted in many different models of 
BSM physics and here we consider the model of universal extra
dimensions (UED) \cite{antoniadis1} where all the quark flavours of the 
SM have heavy Kaluza-Klein (KK) excitations. These heavy KK states will 
modify the form factors which in turn will affect the partial width 
$\G (H\to gg)$. 
The UED model, in its simplest form \cite{appel}, has all the SM particles
propagating in a single extra dimension, which is
compactified on an $S_1/Z_2$ orbifold with $R$ as the radius of
compactification. The KK tower resulting on the four dimensional 
space-time has a tree level mass given by
\begin{equation}
 m_n^2 = m^2 + \frac{n^2}{R^2}
\end{equation}
where $n$ denotes the $n^{th}$-level of the KK tower and $m$
corresponds to the mass of the SM particle in question. 

Since the $H \to gg$ proceeds through diagrams containing fermion
triangle loops and the coupling is proportional to the zero-mode mass of the
fermion even in the case of UED, we consider contributions of the KK
tower of the top quark only. The partial decay width for $H \to gg$ with 
SM and UED contribution is \cite{petriello},
\begin{equation} 
\G(H\to gg) = \frac{G_F~m_H^3}{36\sqrt{2}\pi}
\left(\frac{\alpha_s(m_H)}{\pi}\right)^2~|I_q+\sum_n {\tilde I}_{t^{(n)}}|^2
\end{equation} 
where $G_F$ is the Fermi constant, $\alpha_s(m_H)$ is the running QCD
coupling evaluated at $m_H$ and $I_g=\sum_{q} I_q$, $I_q$ being the 
contributions of the loop integrals involving the different 
quark flavors and ${\tilde I}_{t^{(n)}}$ are the additional
contributions of the loop integrals for the UED case. 
These functions are given in ref~\cite{petriello}. 
The UED contribution is summed over the first few levels of the KK tower, 
till the effects of the higher modes decouple and hardly contribute to 
the amplitude anymore.

\section{Channel of interest} 
We consider the case of a 500~GeV linear $e^+e^-$
collider and calculate the production of a Higgs in association with
a $Z$ boson through a process of the form
\begin{equation} 
e^+e^- \to Z + H \nonumber 
\end{equation} 
where the $Z$ decays leptonically.
In the preceding paragraph we discuss the final states relevant for our
study. 
\begin{enumerate}
\item $e^+ e^- \to \ell^+ \ell^- +$ two jets, which arises when the
Higgs boson decays to a pair of light quarks or gluons\footnote{We
exclude $H \to \tau^\pm$ decays because these produce narrow jets which can
be identified as $\tau^\pm$ with 80-90\% efficiency.}, which then
undergo fragmentation to form a pair of hadronic jets. Clearly, for a
Higgs boson in the SM,  final state will receive contributions 
mainly from the decays $H \to b\bar b$ and $H \to c \bar c$, with a 
minuscule contribution due to $H \to gg$. However, due to the increase
in the partial width for $H \to gg$ due to the extra contribution
coming from the additional KK excitations of the top quark in the loops, 
there will be an enhancement in the overall branching ratio to jets.

\item $e^+ e^- \to \ell^+ \ell^- + b \bar b$, which simply means that
the final state in the above contains two tagged $b$-jets. The decay
width for $H \to b \bar b$ is roughly the same in SM as well as UED, 
although the change of the two-gluon decay mode will have a small effect 
on the branching ratio for the $b \bar b$ mode. 
\end{enumerate}
\begin{figure}[htbp]
\centerline{
\epsfxsize=5cm\epsfysize=5cm\epsfbox{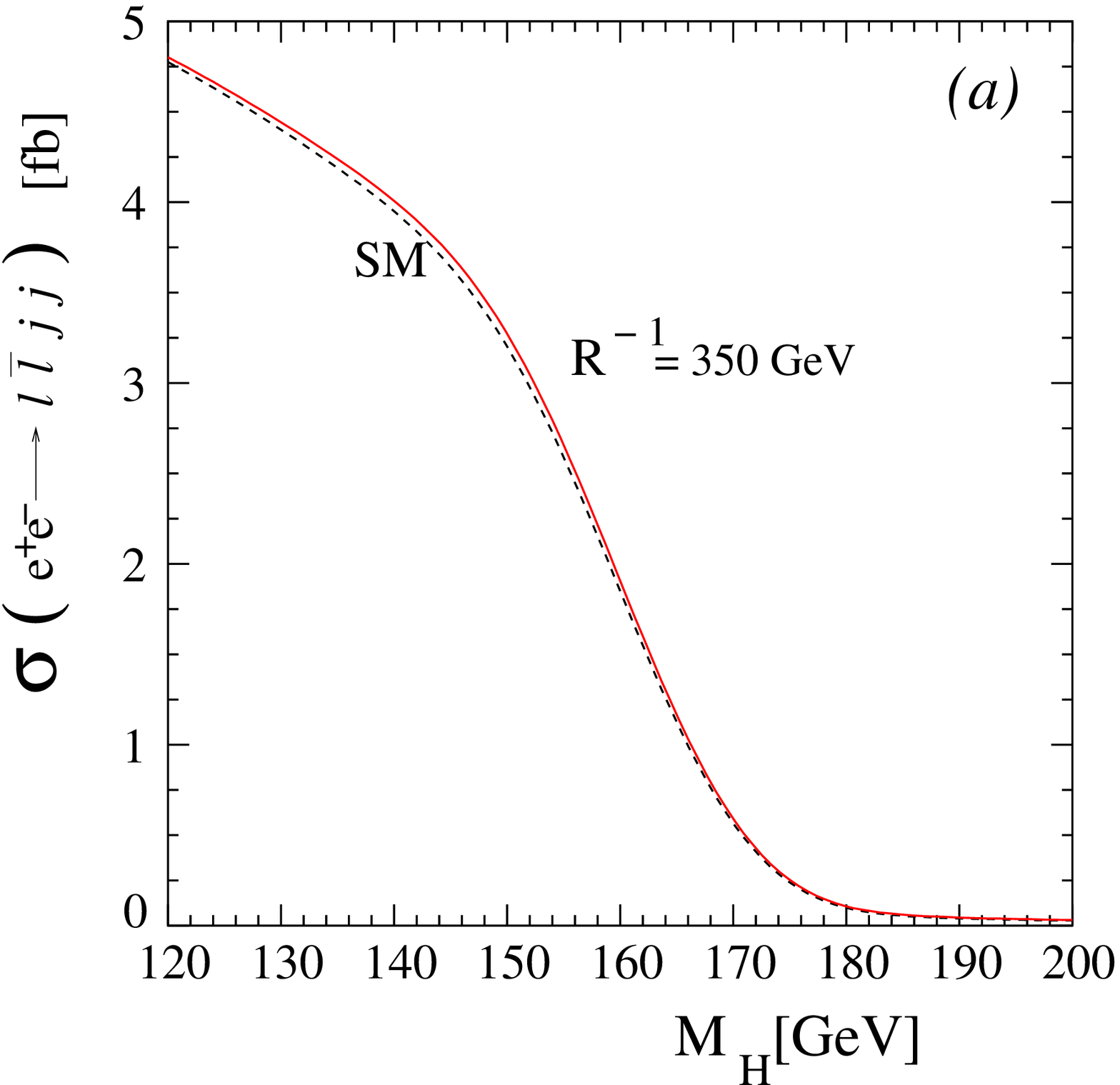}\epsfxsize=5.5cm\epsfbox{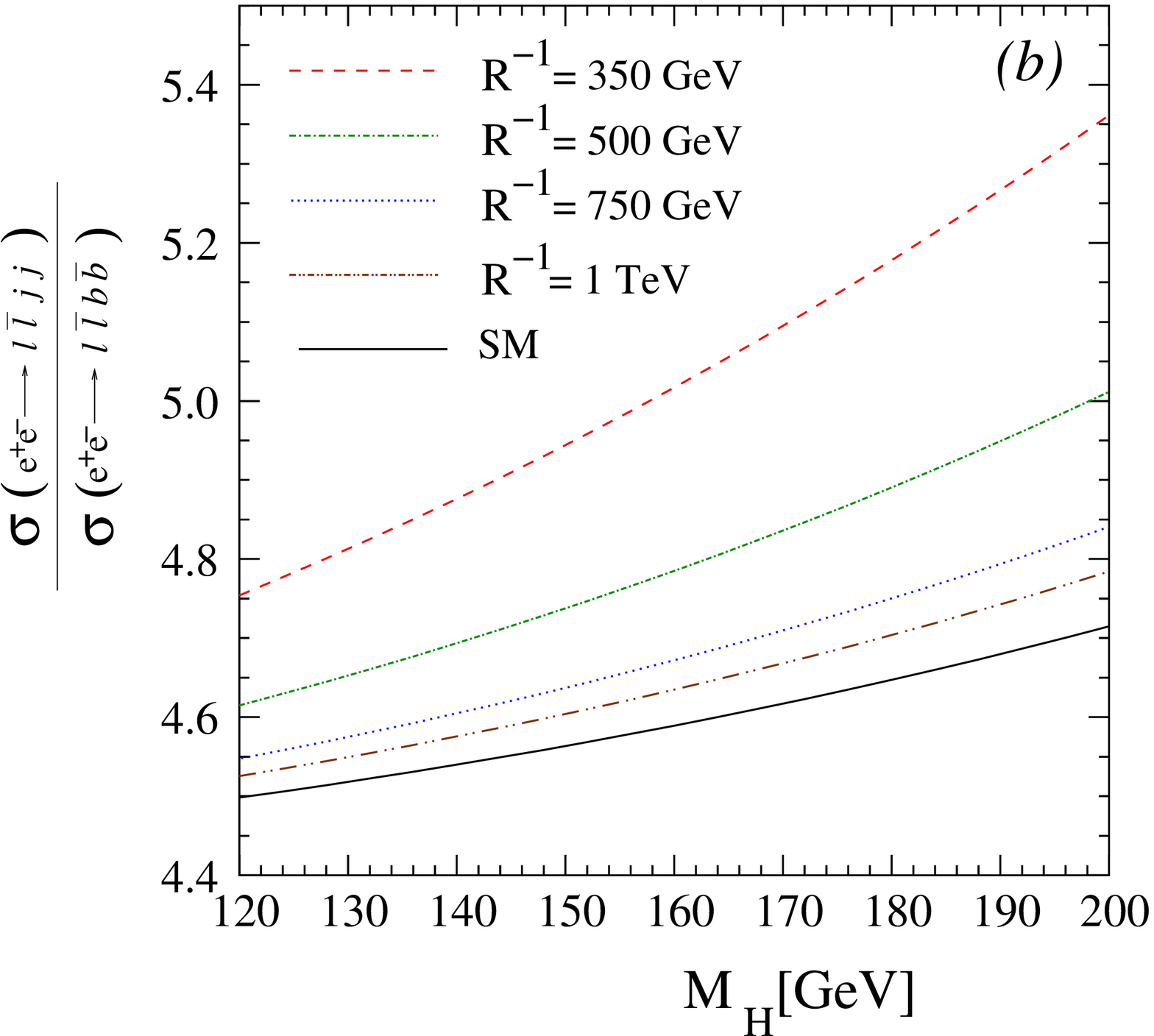}}
\caption{\sl\small The curves are generated for $\sqrt{s}=500$ GeV
linear $e^+e^-$ collider.(a) Shows the cross-section. 
(b) Shows the ratio of cross-sections between two final states for 
different values of the compactification radius.}
\label{fig1}
\end{figure}
\noindent
In our subsequent analysis, we have imposed a few kinematic acceptance 
cuts on the final state particles, viz.,
\begin{enumerate}
\item The final-state leptons should have transverse momentum
$p_T^{(\ell)} > 10$~GeV and pseudo-rapidity $\eta^{(\ell)} < 3.0$.
The final-state jets should have transverse momentum $p_T^{(J)} >
10$~GeV and pseudo-rapidity $\eta^{(J)} < 2.5$.
\item The final-state jets should be clearly separated from each other,
so we impose a cut: $\Delta R_{JJ} (\equiv \sqrt{\Delta \eta_{JJ}^2 +
\Delta \phi_{JJ}^2}) > 0.4$. 
\end{enumerate}

In Figure~\ref{fig1}(a), we illustrate our result for the process discussed
above, namely, 
 $$ e^+e^- \to \ell^+\ell^- ~+~ two ~jets $$
at a $\sqrt{s}=500$ GeV, $e^+e^-$ collider.
The solid (red) line denotes the UED-included cross-section where we
have chosen the value of $R^{-1} = 350$ GeV which gives a greater
enhancement compared to values of $R^{-1}$ greater than the above, while
the dashed (black) line denotes the SM contribution only.  It should
be noted that the graph shows the excess cross-section after removing
the non-Higgs part of the Standard Model contributions (such as
$e^+e^- \to ZZ, ZZ^*$ etc.). The continuum background
($\g^\ast \g^\ast$, $Z^\ast Z^\ast$) too can be easily
neglected as it lies below $10^{-3}~fb$ (in the bins of $b \bar b$
invariant mass) and would hardly affect the rates for the signal in
consideration. The cross-section shown in Figure~\ref{fig1}(a) makes
it clear that it is very hard to see for differences by just looking
at the rates. The cross-sections for $\ell^+\ell^- + {\rm two~jets}$
final state are almost identical in the two cases.  As the
cross-sections look very similar, it would require very precise
measurements to form a distinction between the two cases.
However if we consider the ratio of the two processes,
viz.  $\frac{\s(e^+e^- \to \ell^+\ell^- ~+~ two
  ~jets)}{\s(e^+e^- \to \ell^+\ell^- + b \bar b)}$ we can see that the
difference between the two cases becomes more prominent. In
Figure~\ref{fig1}(b) we plot this ratio for different values of the
compactification scale $R^{-1}$. We find that the ratio differs from
that of the SM throughout the mass range of $120~{\rm GeV} \leq m_H
\leq 200~{\rm GeV}$ with the lines converging towards the SM value as
$R^{-1}$ is increased. In fact this highlights the decoupling
nature of the higher levels of the KK tower and justifies our
termination of the sum of KK towers in the loop, to values where the
contributions become negligibly small. The ratios tend to diverge more
as the Higgs mass increases. This is because the branching ratios of
$H\to gg$ and $H\to b\bar{b}$ become comparable and the enhancement in
the $H\to gg$ mode starts playing a more significant role in the 2-jet
final state. However, there is a caveat.  For comparatively higher
Higgs masses, cross-sections for both the above processes are
small. Hence, we need higher luminosity to differentiate between the
SM and the UED cases in a statistically significant way.  The
robustness of this method is, nevertheless, highlighted in the fact
that although the $H\to gg$ branching ratio is more than an order
smaller than that of $H\to b\bar{b}$ in the intermediate mass range
for the Higgs boson, we are still able to identify the difference due
to the UED contribution which would have been otherwise very difficult
to see, by just looking at the cross-sections.
The ratios are not susceptible to uncertainties like
different efficiency factors associated with particle identifications
as they would cancel out. The efficiency factors will however give a more
realistic estimate of the events that will be observed at the
experiments and which gives us an estimate of the uncertainties in the
statistics.  

In fact the above analysis can be used to identify other new physics
scenarios which play a similar role in modifying the partial width of the
Higgs to massless gauge bosons. It can also be used to distinguish scalars
of other theories which behave similar to the Higgs boson. 
Radions predicted in models of warped extra dimesnsions \cite{RS} have 
similar couplings like the Higgs boson. A major difference is the 
enhanced coupling of Radion to gluons through the trace anomaly. The 
above analysis proves useful in distinguishing Radions from Higgs boson 
quite effectively \cite{SKR}. 

\section{Summary}
This talk was based on the work done with Anindya Datta and further
details can be obtained from Reference~\cite{SKR2}. 

\end{document}